\documentclass{article}
\usepackage{epsfig}

\date{\today}

\begin{document}
\begin{center}
{\Large\bf Nutty dyons}
\vspace{0.5cm}
\\
{\bf Yves Brihaye$^{1}$ }
{\bf and Eugen Radu$^{2}$ }

\vspace*{0.2cm}
{\it $^1$Physique-Math\'ematique, Universite de
Mons-Hainaut, Mons, Belgium}

{\it $^2$ Department of
Mathematical Physics, National University of Ireland Maynooth, Maynooth,
Ireland}
\vspace{0.5cm}
\end{center}
\begin{abstract}
We argue that the Einstein-Yang-Mills-Higgs theory presents
nontrivial solutions with a NUT charge. These solutions approach
asymptotically the Taub-NUT spacetime and generalize the known dyon
black hole configurations. The main properties of the solutions and
the differences with respect to the asymptotically flat case are
discussed. We find that a nonabelian magnetic monopole placed in the field
of gravitational dyon necessarily acquires an electric field, 
while the magnetic charge may take arbitrary values.
\end{abstract}

\section{Introduction}
A feature of certain gauge theories is that they admit classical
solutions which are interpreted as representing magnetic monopoles.
For nonabelian gauge fields interacting with a Higgs scalar, there exist
even regular configurations with a finite mass, as proven by the
famous t'Hooft-Polyakov solution \cite{'tHooft:1974qc}. Typically,
the magnetic monopoles admit also electrically charged
generalizations - so called dyons, the Julia-Zee solution
\cite{Julia:1975ff} of the SU(2)-Higgs theory possibly being the
best known case. 
These solutions admits also gravitating generalizations, both regular and black hole solutions
being considered in the literature  (see
\cite{Volkov:1999cc} for a general review of this topics).
In SU(2)-Einstein-Yang-Mills-Higgs (EYMH) theory, a branch of globally regular gravitating dyons
emerges smoothly from the corresponding flat space solutions.
The  nonabelian black hole solutions emerge from the globally regular configurations,
when a finite regular event horizon radius is imposed \cite{Brihaye:1999nn, Breitenlohner:1991aa}.
These solutions cease to exist beyond some maximal value of the coupling constant 
$\alpha$ (which is proportional to the ratio of the vector meson mass and Planck mass).

It has been speculated that such configurations might have played
an important role in the early stages of the evolution of the
Universe. Also, various analyses indicate that the monopole and dyon
solutions are important in quantum theories.

Since general relativity shares many similarities with gauge
theories, one may ask whether Einstein's equations present solutions
that would be the gravitational analogous of the magnetic monopoles
and dyons. The first example of such a solution was found in
1963 by Newman, Unti and Tamburino (NUT)  \cite{NUT,Misner}. This
metric has become renowned for being "a counterexample to almost
anything" \cite{misner-book} and represents a generalization of the
Schwarzschild vacuum solution \cite{Hawking} (see \cite{Lynden-Bell:1996xj}
for a simple derivation of this metric and historical review).
It is usually interpreted as
describing a gravitational dyon with both ordinary and magnetic
mass \footnote{Note that the Taub-NUT spacetime plays also an important 
role outside general relativity.
For example the asymptotic motion of monopoles in (super-)Yang-Mills theories corresponds to the 
geodesic motion in a Euclideanized Taub-NUT background \cite{Gibbons:1986df}. 
However, these developments are outside the interest of this work.}. 
The NUT charge which plays a dual role to ordinary mass,
in the same way that electric and magnetic charges are dual within
Maxwell theory \cite{dam}.
By continuing the NUT solution through its horizon one arrives in the 
Taub universe \cite{Misner},
which may be interpreted as a homogeneous, 
non-isotropic cosmology with the spatial topology $S^3$.

As discussed by many authors (see e.g. \cite{Hawking:1998jf, Mann:1999pc}),
the presence of magnetic-type mass
(the NUT parameter $n$) introduces a "Dirac-string singularity" in
the metric (but no curvature singularity) . This can be removed by
appropriate identifications and changes in the topology of the
spacetime manifold, which imply a periodic time coordinate.
Moreover, the metric
is not asymptotically flat in the usual sense although it does obey
the required fall-off conditions.

A large number of papers have been written investigating the
properties of the gravitational analogs of magnetic monopoles 
\cite{Gross:1983hb,Sorkin:1983ns}, the
vacuum Taub-NUT solution being generalized in  different directions. 
The corresponding configuration in the Einstein-Maxwell theory 
has has been found in 1964 by Brill \cite{Brill}. 
This abelian solution has been generalized for the matter content of the 
low-energy string theory, a number of NUT-charged configurations being exhibited in the literature
(see e.g. \cite{Johnson:2004zq} for a recent example and a large set of references).
A discussion of the nonabelian counterparts of the  Brill solution
is presented in \cite{Radu:2002hf}. These configurations generalize the well known
SU(2)-Einstein-Yang-Mills hairy black hole solutions \cite{89}, 
presenting, as a new feature, a nontrivial electric potential.
However, the "no global nonabelian
charges" results found for asymptotically flat EYM static
configurations \cite{bizon} are still valid in this case, too. 

Here we present  
arguments for the existence of NUT-charged generalizations of the
known EYMH black hole solutions \cite{Brihaye:1999nn, Breitenlohner:1991aa}. 
 Apart from the interesting
question of finding the properties of a Yang-Mills-Higgs dyon in the field
of a gravitational dyon, there are a number of other reasons to
consider this type of solutions. In some supersymmetric theories,
closure under duality forces us to consider NUT-charged solutions.
Furthermore, dual mass solutions play an important role 
in Euclidean quantum gravity \cite{Hawking:ig} and
therefore cannot be discarded in spite of their causal pathologies.
Also, by considering this type of asymptotics, one may hope to 
attain more general features of gravitating nonabelian dyons.

The paper is structured as follows:  in the next Section
we present the general framework and analyse the field equations and
boundary conditions. In Section 3 we present our numerical results.
We conclude with Section 4, where our results are summarized.

\section{General framework and equations of motion}
\subsection{Action principle}
The action for a  gravitating non-Abelian $SU(2)$ gauge field
coupled to a triplet Higgs field with vanishing Higgs self-coupling is
\begin{eqnarray}
\label{lag0} S=\int \sqrt{-g} d^4x\left ( \frac{R}{16\pi G} -
\frac{1}{2} {\rm Tr} (F_{\mu\nu} F^{\mu\nu}) -\frac{1}{4}{\rm
Tr}(D_\mu \Phi D^\mu \Phi)
  \right ) ,
\end{eqnarray}
with  Newton's constant $G$.  The field strength
tensor is given by
$F_{\mu\nu}=\partial_{\mu}A_{\nu}-\partial_{\nu}A_{\mu}
-ie[A_{\mu},A_{\nu}],$
with
$D_{\mu}=\partial_{\mu}-ie[A_{\mu},\ ]$ being the covariant derivative 
and $e$  the Yang-Mills coupling constant.

 Varying the action
(\ref{lag0}) with respect to $g^{\mu\nu}$, $A_{\mu}$ and $\Phi$ we
have the field equations
\begin{eqnarray}
\nonumber
\label{einstein-eqs} R_{\mu\nu}-\frac{1}{2}g_{\mu\nu}R  &=& 8\pi
G~T_{\mu\nu},
\\
\label{field-eqs}
\frac{1}{\sqrt{-g}}D_{\mu}(\sqrt{-g} F^{\mu \nu}) &=&\frac{1}{4}
ie[\Phi,D^{\nu}\Phi],
\\
\nonumber
\label{Heqs} \frac{1}{\sqrt{-g}}D_{\mu}(\sqrt{-g}D^{\mu} \Phi)&=&0,
\end{eqnarray}
where the stress-energy tensor is
\begin{eqnarray}
T_{\mu\nu} = 2Tr\{F_{\mu \alpha} F_{\nu \beta} g^{\alpha \beta}
-\frac{1}{4}g_{\mu \nu}F_{\alpha \beta}F^{\alpha \beta} \}
+Tr\{\frac{1}{2}D_{\mu}\Phi D_{\nu}\Phi -\frac{1}{4}g_{\mu
\nu}D_{\alpha}\Phi D^{\alpha}\Phi\}.
\end{eqnarray}
\subsection{Metric ansatz and symmetries}
We consider NUT-charged spacetimes whose metric can be written locally in the
form
\begin{eqnarray}
\label{metric} 
ds^2=\frac{dr^2}{N(r)}+P^2(r)(d\theta^{2}+\sin
^2\theta d\varphi^{2}) -N(r)\sigma^2(r)(dt+4 n
\sin^2(\frac{\theta}{2}) d\varphi)^{2},
\end{eqnarray}
the NUT parameter $n$ being defined as usually in terms of the coefficient
appearing in the differential $dt+4 n \sin^2(\theta/2)d\varphi$.
Here $\theta$ and $\varphi$ are the standard angles parametrizing 
an $S^2$ with ranges $0 \leq \theta \leq \pi,~ 0 \leq \varphi \leq 2\pi$.

Apart from the Killing vector $K_0=\partial_{t}$, this line element possesses
three more Killing vectors
characterizing the NUT symmetries
\begin{eqnarray}
\label{Killing}
\nonumber
K_1&=&\sin \varphi \partial_{\theta}
+\cos \varphi \cot \theta  \partial_{\varphi}
+2n\cos\varphi \tan \frac{\theta}{2}  \partial_{t},
\\
K_2&=&\cos \varphi \partial_{\theta}
-\sin \varphi \cot \theta\partial_{\varphi}
-2n\sin \varphi\tan \frac{\theta}{2}\partial_{t},
\\
\nonumber
K_3&=&\partial_{\varphi}-2n\partial_{t}.
\end{eqnarray}
These  Killing vectors form a subgroup with the same structure constants
that are obeyed by spherically symmetric
solutions $[K_a,K_b]=\epsilon_{abc}K_c$.

The $n\sin^2 (\theta/2)$ term in the metric means that a small loop around the
$z-$axis does not shrink to zero at $\theta=\pi$.
This singularity can be regarded ar the analogue of a Dirac string in electrodynamics
 and is not related to the usual degeneracies of spherical coordinates on the two-sphere.
This problem was first encountered in the vacuum NUT metric.
One way to deal with this singularity has been proposed by Misner \cite{misner-book}.
His argument holds also independently of the precise functional form of $N$ and $\sigma$.
In this construction, one considers one coordinate patch in which the string runs off to
infinity along the north axis.
A new coordinate system can then be
found with the string running off to infinity along the south axis
with $t'=t+4n\varphi,$ the string becoming an artifact resulting 
from a poor choice of coordinates.
It is clear that the 
$t$ coordinate is also periodic with period
$8 \pi n$ and essentially becomes an Euler angle coordinate on $S^3$.
Thus an observer with $(r,\theta,\varphi)=const.$ follows a closed timelike curve.
These lines cannot be removed by going to a covering space
and there are no reasonable spacelike surface.
One finds also that surfaces of constant radius have the topology
of a three-sphere, in which there is a Hopf fibration of the $S^1$
of time over the spatial $S^2$ \cite{misner-book}.

Therefore for $n$ different from zero, the  metric structure
(\ref{metric}) generically shares the same troubles exhibited by the vacuum
Taub-NUT gravitational field \cite{Mueller:1986ij}, and the solutions cannot be
interpreted properly as black holes.

\subsection{Matter fields ansatz}
While the  Higgs field is given by the usual form
\begin{eqnarray}
\Phi=\phi \tau_3,
\end{eqnarray}
the computation of the appropriate $SU(2)$ connection compatible
with the Killing symmetries (\ref{Killing})  is a more involved
task. This can be done by applying the standard rule for calculating
the gauge potentials for any spacetime group
\cite{Forgacs:1980zs,Bergmann:fi}. According to Forgacs and Manton,
a gauge field admit a spacetime symmetry if the spacetime
transformation of the potential can be compensated by a gauge
transformation \cite{Forgacs:1980zs}
$ {\mathcal{L}}_{K_i} A_{\mu}=D_{\mu}W_{i},$
 where ${\mathcal{L}}$ stands for the Lie derivative.

Taking into account the symmetries of the line element (\ref{metric})
we find the general form
\begin{eqnarray}
\label{A} A=\frac{1}{2e} \Big \{  \Big(dt+4n \sin
^2(\frac{\theta}{2}) d\varphi\Big)u(r) \tau_3+ \nu(r) \tau_3 dr+
\Big(\omega(r) \tau_1 +\tilde{\omega}(r) \tau_2\Big) d \theta
\\
\nonumber
 +\Big[\cos \theta  \tau_3+ (\omega(r)
\tau_2-\tilde{\omega}(r)\tau_1 ) \sin \theta  \Big]d \varphi \Big\}.
\end{eqnarray}
This gauge connection remains invariant under a residual $U(1)$
gauge symmetry which can be used to set $\nu=0$. Also, because the
variables $\omega$ and $\tilde{\omega}$ appear completely
symmetrically in the EYMH system, the two amplitudes must be
proportional and we can always set $\tilde{\omega}=0$ (after a
suitable gauge transformation).  Thus, similar to the $n=0$ case, 
the gauge potential is described by two functions
$\omega(r)$ and $u(r)$ which we shall refer to as magnetic and
electric potential, respectively.

\subsection{Field equations and known solutions}

 Within the above ansatz, the classical equations of motion
can  be derived from the following reduced action
\begin{eqnarray}
\label{action} S = \int  ~dr~dt \Big [ \frac{1}{8 \pi G}\Big (\sigma(1-NP'^2-PP'N')+
2P'(\sigma NP)'+\frac{n^2\sigma^3N}{P^2} \Big)
\\
\nonumber
-\Big( \frac{1}{e^2}( N\sigma \omega'^2+
\frac{\sigma(\omega^2-1+2nu)^2}{2P^2}-\frac{P^2 u'^2}{2\sigma^2}
-\frac{\omega^2 u^2}{\sigma N})+\frac{1}{2}\sigma NP^2 \phi'^2+\sigma \omega^2 \phi^2\Big) 
\Big],
\end{eqnarray}
where the prime denotes the derivative with respect to the radial
coordinate $r$.

At this point, we fix the metric gauge by choosing $P(r)=\sqrt{r^2+n^2}$, which allows
a straightforward analysis of the relation with the abelian configurations.

Dimensionless quantities are obtained by considering the rescalings
$r\to r/(\eta e)$, $\phi \to \phi\eta$, $n \to n/(\eta e)$, $u \to
 \eta e u$ (where $\eta$ is the asymptotic magnitude of the Higgs field). 
As a result, the field equations depend only on the
coupling constant $\alpha  = \sqrt{4\pi G}\eta$.

The EYMH equations  reduce to the following system of five
non-linear differential equations
\begin{eqnarray}
\nonumber
rN'&=&1-N + \frac{n^2 N}{P^2}(3 \sigma^2-1)-2 \alpha^2 \big(N \omega'^2+
\frac{1}{2P^2}(\omega^2-1+2nu)^2
\\
\nonumber
&&~~~~~+\frac{P^2 u'^2}{2\sigma^2}+\frac{\omega^2 u^2}{\sigma^2 N}
+\frac{1}{2}NP^2 \phi'^2+\omega^2 \phi^2 \big),
\\
\nonumber 
\sigma'&=&\frac{n^2 \sigma(1-\sigma^2)}{r P^2}+\frac{\alpha^2 \sigma}{r} (P^2 \phi'^2+2 \omega'^2+\frac{2 \omega^2 u^2}{\sigma^2 N^2}),
\\ 
\label{e4}
(N\sigma \omega')'&=&\sigma \omega \big(\frac{(\omega^2-1+2nu)}{P^2}+\phi^2-\frac{u^2}{\sigma^2 N} \big),
\\
\nonumber
(N\sigma P^2 \phi')'&=&2 \sigma \omega^2 \phi,
\\
\nonumber
\big(\frac{P^2 u'}{\sigma}\big)'&=&\frac{2 \omega^2 u}{\sigma N}-\frac{2n \sigma}{P^2}(\omega^2-1+2nu).
\end{eqnarray}
Two explicit solution of the above equations are well known. 
The vacuum Taub-NUT one corresponds to
\begin{eqnarray}
\label{vacuum}
\omega(r)=\pm 1,~~~~~u(r)=0,~~~~\sigma(r)=1,~~~\phi(r)=1,~~
N(r)=1-\frac{2(Mr+n^2)}{r^2+n^2}.
\end{eqnarray}
The  U(1) Brill solution \cite{Brill}  has the form 
\begin{eqnarray}
\label{Brill}
\omega(r)=0,~~u(r)=u_0+\frac{nQ_m-Q_er}{r^2+n^2},~\sigma(r)=1,~\phi(r)=1,~~
N(r)=1-\frac{2(Mr+n^2)}{r^2+n^2}+\frac{\alpha^2(Q_e^2+Q_m^2)}{4(r^2+n^2)}.
\end{eqnarray}
and describes a gravitating abelian dyon with a mass $M$, 
electric charge $Q_e$ and magnetic charge $Q_m\equiv 1-2u_0n$, $u_0$ being an arbitrary constant,
corresponding to the asymptotic value of the electric potential.

It can be stressed that the Brill solution possesses two, one
or zero horizons, according to the values of the free parameters
$Q_e, M, u(\infty)$.
In the same way as in the case of Reissner-Nordstr\"om
solutions, the extremal Brill solution can be defined
as the solutions with a degenerate horizon at $r=r_0$.
This gives the following conditions, fixing $M$ and $r_0$
\begin{equation}
   r_0 = M \ \ , \ \ M^2 + n^2 - \frac{\alpha^2}{4}(Q_e^2 + Q_m^2) = 0.
\end{equation}
As we will see later, it is convenient to further specify the
arbitrary constant $u(\infty)$ in such a way the $u(r_0)=0$,
this implying
\begin{equation}
   \frac{1-Q_m}{2n} + \frac{n Q_m - M Q_e}{M^2+n^2}=0,
\end{equation}
which fixes $Q_m$ and leaves $Q_e$ as the only remaining free parameter.
In the following we will refer to this solution as to the
extremal Brill solution. As far as we could see, it is  not
possible to express $M$ and $Q_m$ in a closed form depending
on $(\alpha,~n,~Q_e)$, but the solution can be constructed numerically.

\subsection{Boundary conditions}
We want the metric (\ref{metric}) to describe a nonsingular,
asymptotically NUT spacetime outside an horizon located at $r=r_h$.
Here $N(r_h)=0$ is only a coordinate singularity where all curvature
invariants are finite. A nonsingular extension across this null
surface can be found just as at the event horizon of a black hole.
If the time is chosen to be periodic, as discussed above, this
surface would not be a global event horizon, although it would still
be an apparent horizon. The regularity assumption implies that all
curvature invariants at $r=r_h$ are finite.

The corresponding expansion as $r \to r_h$ is
\begin{eqnarray}
\label{eh} 
\nonumber 
N(r)&=&N_1(r-r_h)+O(r-r_h)^2,
\\
\nonumber 
\sigma(r)&=&\sigma_h+\sigma_1(r-r_h)+O(r-r_h)^2,
\\
\omega(r)&=&\omega_h+\omega_1(r-r_h)+O(r-r_h)^2,
\\
\nonumber 
u(r)&=&u_1(r-r_h)+u_2(r-r_h)^2+O(r-r_h)^3,
\\
\nonumber \phi(r)&=&\phi_h+\phi_1(r-r_h)+O(r-r_h)^2,
\end{eqnarray}
where $P_h^2=r_h^2+n^2$ and
\begin{eqnarray}
\nonumber
N_1&=&\frac{1}{r_h}\left( 1- 2 \alpha^2 
(\frac{(\omega_h^2-1)^2}{2P_h^2}+\frac{1}{2} \frac{u_1^2 P_h^2}{\sigma_h^2}
+\omega_h\phi_h^2  \right),~~ 
\sigma_1=
\frac{n^2 \sigma_h(1-\sigma_h^2)}{r_h P_h^2}+\frac{\alpha^2 \sigma_h}{r_h} (P_h^2 \phi_1^2+2 \omega_1^2+\frac{2 \omega_1^2 u_1^2}{\sigma_h^2 N_1^2}),
\\
\omega_1&=&\frac{ \omega_h}{N_1} \left(\frac{ \omega_h^2-1}{P_h^2}+ \phi_h^2 \right),
~~
u_2= \frac{\sigma_1 u_1}{2 \sigma_h}-\frac{n \sigma_h^2(\omega_h^2-1)}{P_h^4}
+\frac{u_1\omega_h^2}{N_1P_h^2}-\frac{u_1r_h }{P_h^2},
~~ \phi_1=\frac{2\omega_h^2\phi_h }{N_1 P_h^2},
\end{eqnarray}
 $\sigma_h,~u_1,~\omega_h,~\phi_h$ being arbitrary parameters.

The analysis of the field equations as $r\to\infty$ gives the
following expression in terms of the constants $c,~u_0,~Q_e,~\tilde
\phi_1,~M$
\begin{eqnarray}
\label{asimpt}
\nonumber
 N(r)  &\sim&   1-\frac{2M}{r}-\frac{2n^2-\alpha^2\left(\tilde\phi_1^2+(1-2nu_0^2)^2+Q_e^2\right)}{r^2}
 +\frac{M(2n^2+\alpha^2 \tilde\phi_1^2)}{r^3}+\dots,
\\
\sigma &\sim& 1-\frac{\alpha^2  \tilde\phi_1^2}{2 r^2}- \frac{4\alpha^2  \tilde\phi_1^2 M}{3r^3} +\dots,
~~~\omega(r)  \sim  c
e^{-\sqrt{1-u_0^2}r}+\dots,
\\
\nonumber
\phi  &\sim& 1-\frac{\tilde\phi_1}{r}+\frac{\tilde\phi_1 M}{r^2}+\dots,
~~u(r)\sim u_0-\frac{Q_e}{r}+\frac{n(1-2n u_0)}{r^2}-\frac{Q_e (6n^2+\alpha^2\tilde\phi_1^2)}{6r^3}+\dots
\end{eqnarray}
Note that similar to the
$n=0$ asymptotically flat case, the magnitude of the electric potential at infinity
cannot exceed that of the Higgs field, $|u_0|<1$
\footnote {This depends on the asymptotic structure of the spacetime. 
For example, in an anti-de Sitter spacetime,  $u_0$ may take arbitrary 
values \cite{vanderBij:2002sq}.}. The constant $M$ appearing
in the asymptotic expansion of the metric function $N(r)$ can be
interpreted as the total mass of solutions (this can be proven rigurously by applying the
general formalism proposed in \cite{Magnon}). Note that $M$ and $n$ are unrelated
on a classical level.

Also,  no  purely monopole
solution can exist for a nonvanishing NUT charge 
(i.e. one cannot consistently set $u=0$ unless
$\omega=\pm 1$, in which case  the vacuum Taub-NUT solution is
recovered). Thus, a nonabelian magnetic monopole placed in the field
of gravitational dyon necessarily acquires an electric field.

We close this section by remarking that the definition of the
nonabelian charges   is less clear for $n \neq 0$. Although we may
still define a 't Hooft field strength tensor, in the absence of 
a nontrivial two-sphere at infinity on which to integrate,
the only reasonable
definition the nonabelian magnetic and electric charges is in terms
of the asymptotic behavior of the gauge field. By analogy to the  
asymptotically flat case, $Q_e$ and $Q_m$ are defined from
$F_{tr}^{(3)} \simeq Q_e/r^2$ and $F_{\theta \phi}^{(3)} \simeq Q_m
\sin \theta $ (a similar problem occurs for an $U(1)$ field
\cite{Johnson:1994ek}).
Thus, since $Q_m=1-2nu_0$, the usual quantization 
relation for the magnetic charge is lost for $n\neq 0$, 
which is a consequence of the pathological large scale structure of a NUT-charged spacetime. 

\newpage
\setlength{\unitlength}{1cm}

\begin{picture}(18,7)
\centering
\put(2,0.0){\epsfig{file=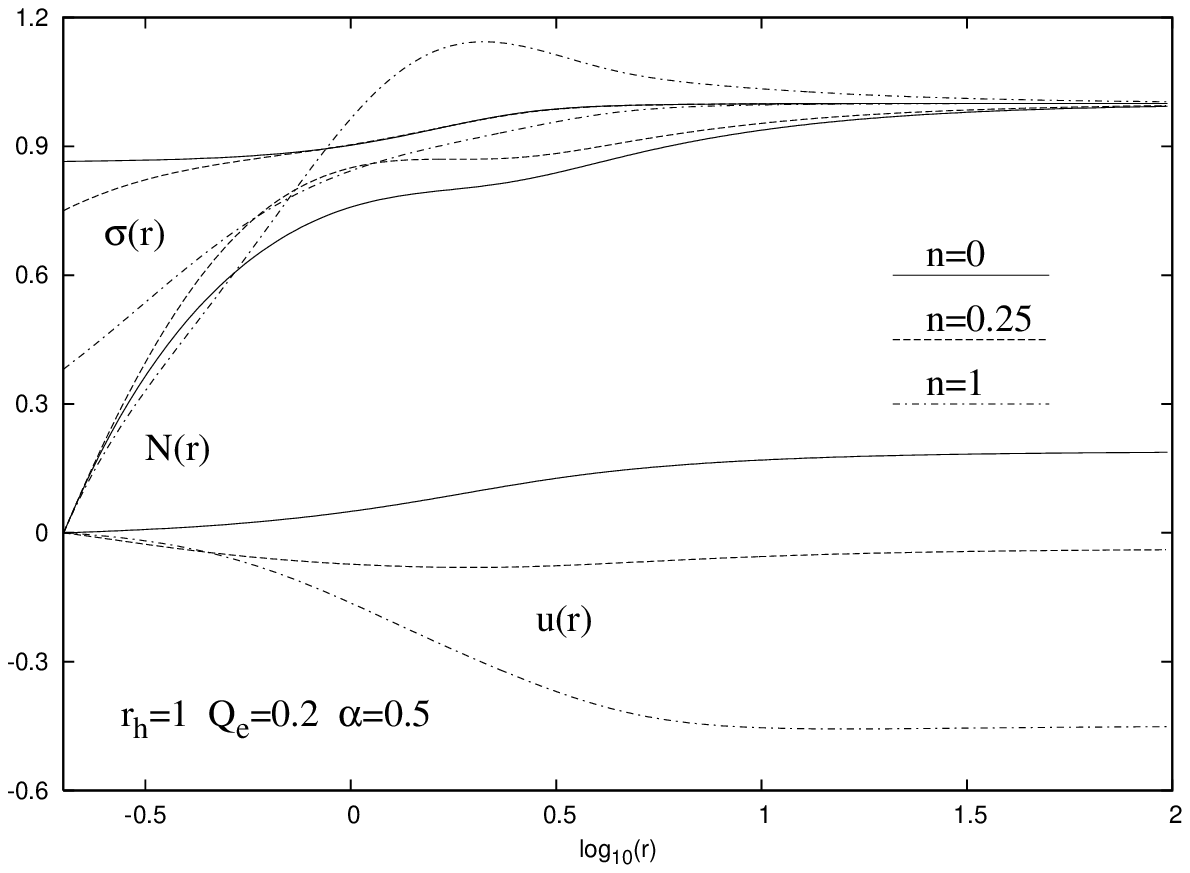,width=12cm}}
\end{picture}
\\
\\
{\small {\bf Figure 1.}
The functions $N(r),~\sigma(r)$ and $u(r)$ are plotted
for three typical solutions for the same values of  $(r_h,~Q_e,~\alpha$)}.
\\
\\

\section{Numerical results}
Although an analytic or approximate solution appears to be
intractable, we present in this section numerical arguments that the
known EYMH black hole solutions can be extended to include a NUT
parameter.

The equations of motion (\ref{e4}) have been solved for a large set of
the parameters ($\alpha,~n,~Q_e,~r_h$), 
looking for solutions interpolating between the asymptotics (\ref{eh}) and (\ref{asimpt}).
NUT-charged solutions are found for any $n=0$ EYMH dyonic black hole 
configuration by slowly increasing the
parameter $n$ 
(since the transformation $n\to -n$ leaves the field equations unchanged  
except for the sign of the electric potential,
we consider here only positive values of $n$).
As expected, these configurations have many features in
common with the $n=0$  solutions
discussed in \cite{Brihaye:1999nn}; they also present new features that
we will  pointed out in the discussion.
Typical profiles for the metric functions $N(r)$ and $\sigma(r)$ and for the electric potential
$u(r)$ are presented in Figure 1, for a dyonic black hole solution as well 
as for two NUT-charged solutions.
The gauge function $\omega(r)$ and the Higgs scalar $\phi(r)$ interpolates monotonically 
between some constant values on the event horizon and zero respectively one at infinity, 
without presenting any local extremum (see Figure 3).

The domain of existence of the nonabelian nutty dyons
can be determined in the space of  parameters.
If we fix the electric charge $Q_e$ of the solution, then 
there likely exist a volume ${\cal V}_Q$ in the parameter space of
$(\alpha, n, r_h)$ inside which nonabelian solutions exist 
and on the side of which they become singular and/or bifurcate
into abelian solution of the type of the Brill solution. 
For $n=0$ the domain of the
$(\alpha,~r_h)$ plane where nonabelian solutions exist was determined in
\cite{Breitenlohner:1991aa} for $Q_e=0$ and in \cite{Brihaye:1999nn} for $Q_e\neq 0$.

The determination of ${\cal V}_Q$ is of course a huge task.
 In this letter, we will not attempt to determine
the shape of  ${\cal V}_Q$ accurately but rather attempt to sketch it
 by analyzing the pattern of solutions on some generic lines in the
 space of parameters. 
 For definiteness we set $Q_e=0.2$ in our numerical analysis, although nontrivial 
solutions have been found also for other values of the electric charge.

\subsection{$n$ varying}

First, we have integrated the system of equations (\ref{e4}) with fixed values for
$\alpha, r_h$ and $Q_e$ and increased the
NUT charge $n$.
Our values here are $\alpha = 1.0$, $r_h=0.2$ and $Q_e = 0.2$
corresponding
to a generic values for 
\newpage
\setlength{\unitlength}{1cm}

\begin{picture}(18,7)
\centering
\put(2,0.0){\epsfig{file=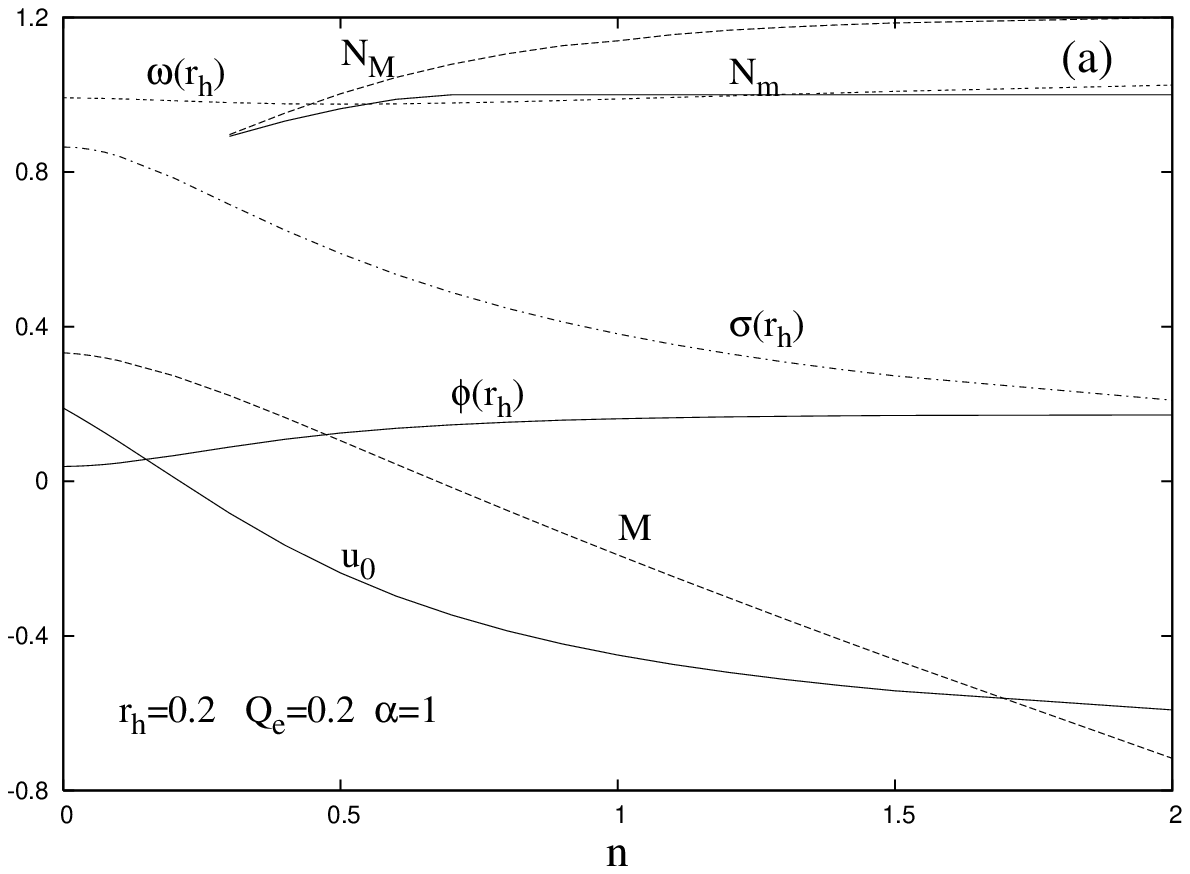,width=12cm}}
\end{picture}
\begin{picture}(19,8.5)
\centering
\put(2.6,0.0){\epsfig{file=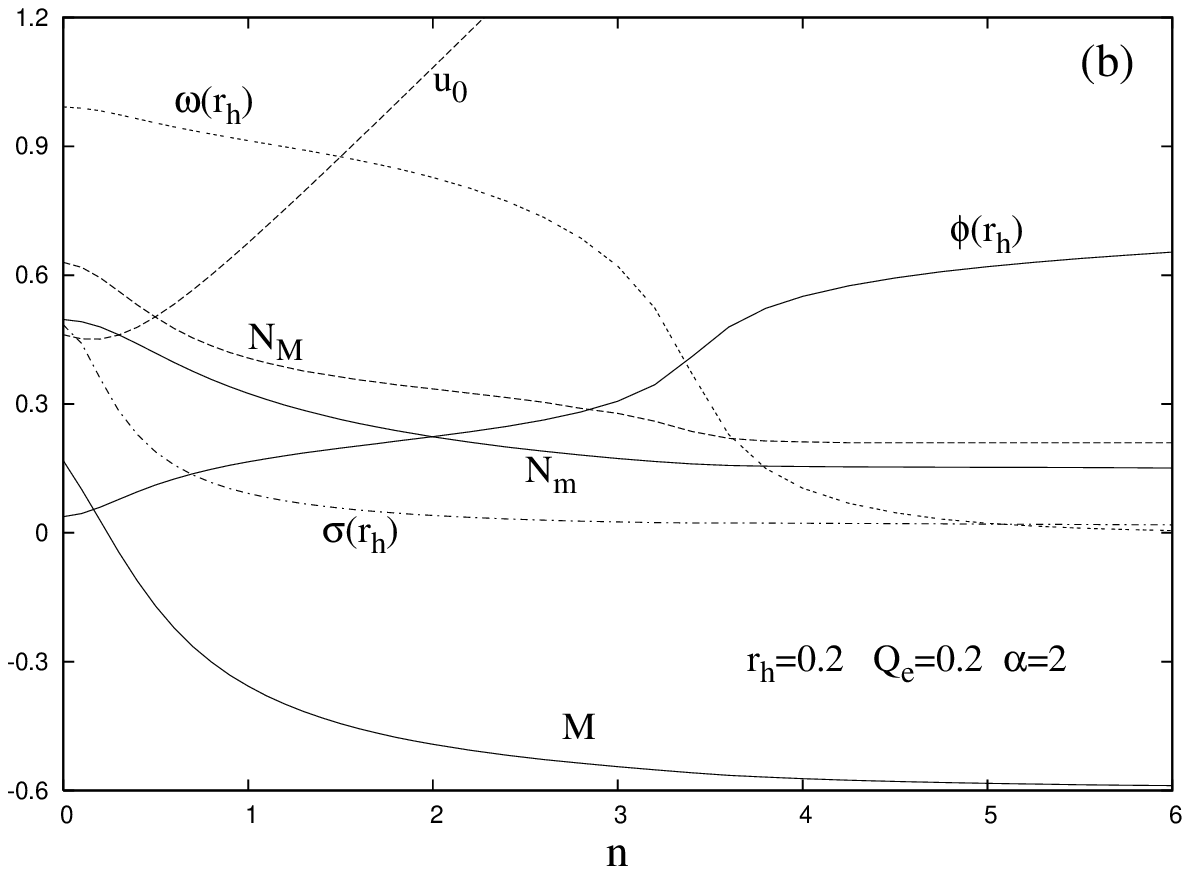,width=12cm}}
\end{picture}
\\
\\
{\small {\bf Figure 2.}
The values  of the parameters $M,~\sigma(r_h),~N_m,~N_M,~\phi(r_h),~\omega(r_h)$ and
$u(\infty)=u_0$ are shown as a function of
$n$  for  solutions with $r_h=0.2,~Q_e=0.2$ and two 
different values of the coupling constant $\alpha$.}
\vspace{0.5cm}
\\
the parameters (the corresponding $n=0$ gravitating
dyon was constructed in \cite{Brihaye:1999nn}).  
As far as the function $u(r)$ is concerned, there exists a main difference
between the case $n=0$ and $n \neq 0$. Indeed in the case  $n=0$ this function
behave asymptotically like  $u(r) \sim u_0 + Q_e/r + O(1/r^4)$  while in the 
presence of  a NUT charge the behaviour is instead   
$u(r) \sim u_0 + Q_e/r + K/r^2$, where, as seen from (\ref{asimpt}),
the constant $K$ increases with $n$. 
Thus, when $n$ becomes large, it becomes more 
difficult to construct numerical solutions with a good enough accuracy
\footnote{To integrate the equations, we used the differential
equation solver COLSYS which involves a Newton-Raphson method
\cite{COLSYS}.}, for a given value of the electric charge $Q_e$.

The effect of increasing $n$ apparently depend strongly of the value $\alpha$.
For $\alpha$ small (typically $\alpha \leq 1$)
the pattern can be summarized by the following points~:
(i) No local extrema of $N(r)$ are found for small enough
\newpage
\setlength{\unitlength}{1cm}

\begin{picture}(18,7)
\centering
\put(2,0.0){\epsfig{file=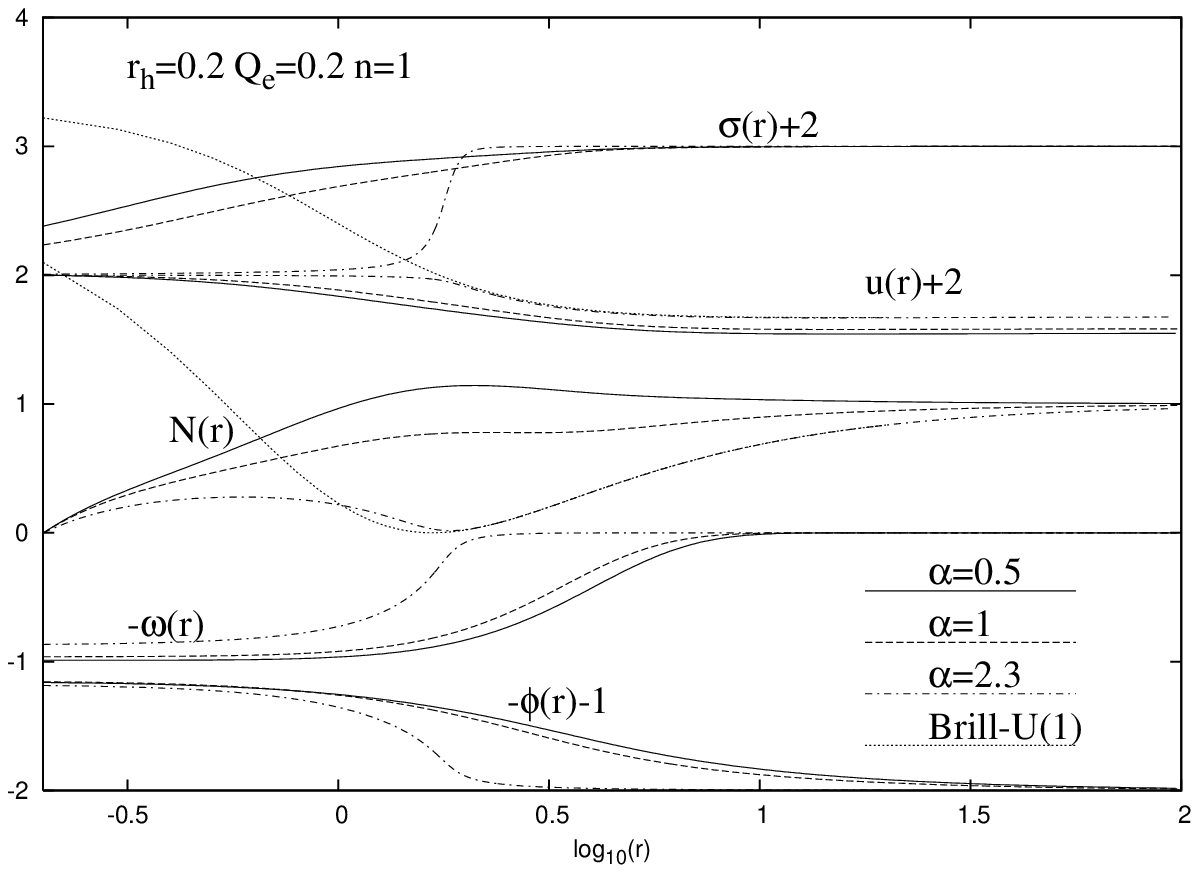,width=12cm}}
\end{picture}
\\
\\
{\small {\bf Figure 3.}
The metric functions $N(r),~\sigma(r)$ and the matter functions $\omega(r),~\phi(r)$ 
and $u(r)$ are shown as a function of $r$  for fixed values of $(r_h,~Q_e,~n$) and 
three different values of $\alpha$.
The functions $N(r)$ and $u(r)$ of the corresponding extremal Brill solution 
(with $\omega(r)=0,~h(r)=1$) are also exhibited.}
\\
\\ 
values of $n$.
For larger $n$, the function $N(r)$ develops a 
local maximum and also a local minimum, say
$N_{M}$ and $N_m$ at some
intermediate, $n-$depending values of $r$. For $n$ large enough, we have
$N_{max} > 1$. No local minimum of $N$ persist for large enough $n$,
the minimum of $N(r)$ ($N_m=1$) being attained as $r\to \infty$.
(ii) The second metric function $\sigma(r)$ still remains
monotonically increasing but the value $\sigma(r_h)$ diminishes when
$n$ increases.
(iii) The asymptotic value $u(\infty)$ also decreases for increasing $n$.
With the values choosen, we have $u(\infty)\approx 0.189$ for $n=0$;
we find $u(\infty)=0$ for $n\approx 0.25$ and
 negative values for larger $n$.

These effects are illustrated on Figure 2a. On this Figure we have set
$0 < n < 2$ but we noticed no significant change of the behaviour
for larger values of $n$.  

For larger values of $\alpha$ (typically $\alpha \geq 2$)
the situation is quite different, namely:
(i) The function $N(r)$  posseses both a local minimum and a local maximum.
(ii) The values $w(r_h)$ and $\phi(r_h)$ increase with $n$ and 
approach respectively zero and one,
suggesting that the solution approaches an Abelian Brill solution.
These results are summarize on Figure 2b for $\alpha = 2$. However, due to 
numerical difficulties, we could not determine properly the value of
$n$ where the bifurcation occur.
The statement of a bifurcation into a Brill solution
is confirmed in the next subsection where $\alpha$ is varying.

Nevertheless, it seems that there are two possible patterns for $n \rightarrow \infty$~:
for 
values of $\alpha$ smaller than a critical value $\hat \alpha$, 
solutions with large values of $n$ seem to occur,
while for $\alpha > \hat \alpha$, the solutions bifurcate into a Brill solution
(for $Q_e = 0.2$ we find $\hat
     \alpha \sim 1.5$).
The occurence of these two patterns is reminiscent to the case of $n=0$ gravitating
dyons.

Note also that, as shown in these plots, the mass parameter $M$ 
takes negative values for large enough values of $n$. 
This is not a surprise, since something similar happens already 
 for the U(1) Brill solution (\ref{Brill}).

\subsection{$\alpha$ varying}
We now discuss the behaviour of the solutions
for a varying $\alpha$  and the other parameters fixed.
In absence of a NUT charge it is know 
\cite{Breitenlohner:1991aa, Brihaye:1999nn} that nonabelian dyonic black hole
exist for $r_h \in [0, \sqrt{3+4Q_e^2}/2]$. For fixed $Q_e$ and $r_h$
and increasing $\alpha$ they
bifurcates into an extremal  Reissner-Nordstr\"om solution at 
$\alpha \sim \alpha_c$. The value $\alpha_c$ depends of course
on $r_h$ and $Q_e$. For $r_h \ll 1$ the value $\alpha_c \approx 1.4$
is found numerically and depends weakly on $Q_e$.
For $r_h \sim \sqrt{3+4Q_e^2}/2$ 
we have $\alpha_c \approx \sqrt{(3+4Q_e^2)/(1+Q_e^2)}/2$.

For $n>0$ we see (e.g. on Figure 3) that the local maximum
characterizing the function $N(r)$ of a
nutty solution (at least for large enough
values of the NUT charge $n$)
progressively disappears in favor of a local
\newpage
\setlength{\unitlength}{1cm}

\begin{picture}(18,7)
\centering
\put(2,0.0){\epsfig{file=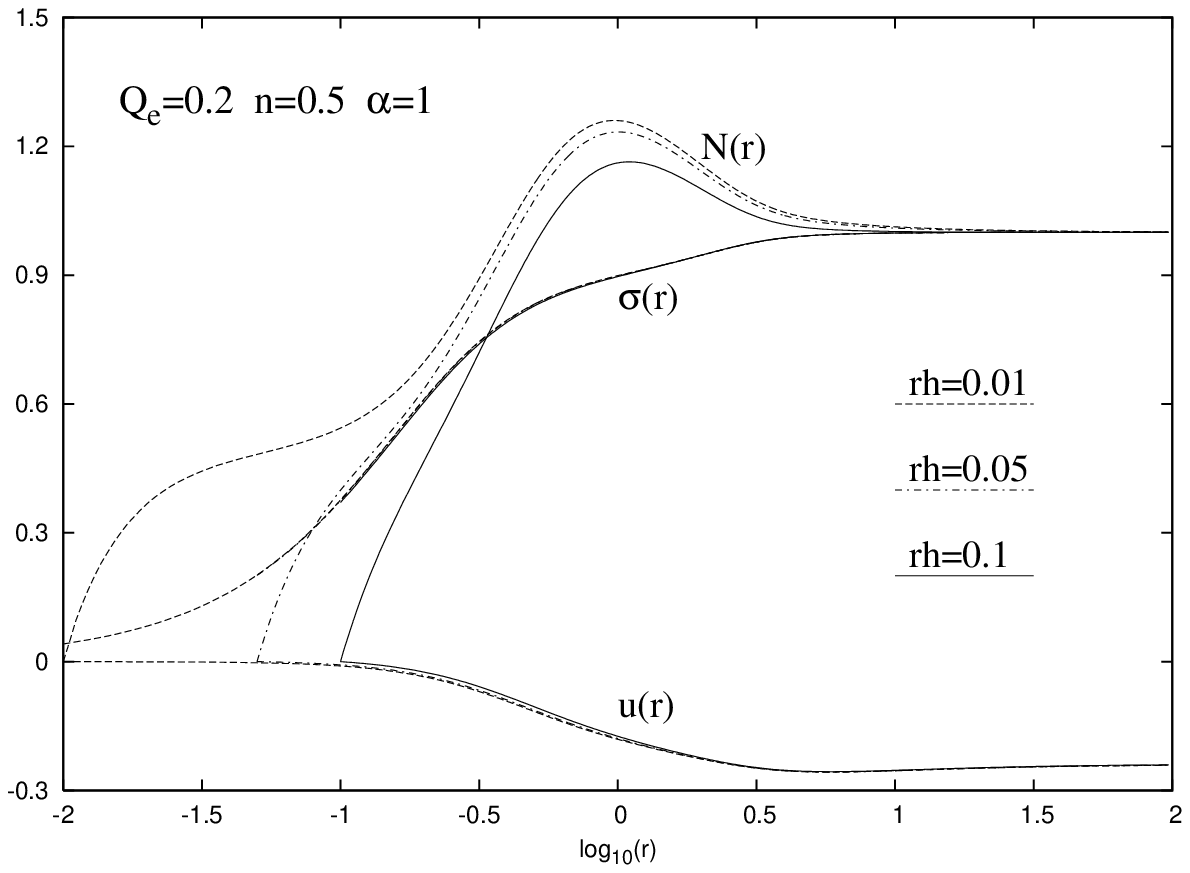,width=12cm}}
\end{picture}
\\
\\
{\small {\bf Figure 4.}
The profiles of the functions $N(r),~\sigma(r)$ and $u(r)$ are represented for $Q_e=0.2,~n=0.5,~\alpha=1$ and three different
values of $r_h$ .}
\\
\\
minimum when $\alpha$ increases. This minimum
appears far outside the event horizon $r=r_h$
and becomes deeper. In fact,
the minimal value $N_{m}$ approaches zero when $\alpha$
tends to a critical value, say $\alpha_c(n,Q_e,r_h)$. If we denote
by $r_m$ the value of the radial variable where $N(r_m)=0$
(with $r_m > r_h$) our numerical results strongly indicate
that the non abelian solution converges into an extremal Brill solution
on the interval
$r \in [r_m, \infty)$ for $\alpha \rightarrow \alpha_c$.

Indeed, the matter functions's profiles $u,~w,~\phi$ and the metric
functions $\sigma,~N$ all approaches the profiles of the
corresponding extremal Brill solution with the same
$\alpha_c,~Q_e,~n$.  This result is illustrated   on Figure 3
for $n=1,~r_h=0.2$ and $Q_e=0.2$; 
in this case, we find $\alpha_c \approx 2.35$
but we believe that the result holds for generic values of $(n,~r_h,~Q_e)$.
The determination of the critical value $\alpha_c(n,~r_h,~Q_e)$ is not
aimed in this letter. 
However, it seems that the value $\alpha_c$ depends weakly of $n$, for example
we find $\alpha_c \approx 2.22$ for $n \in [3,4]$.

Nevertheless we can conclude that nutty
dyons exist on a finite interval of $\alpha$ and bifurcate into
extremal Brill solutions for $\alpha = \alpha_c$.


\subsection{$r_h$ varying}
In the case of gravitating dyonic black holes with event horizon $r_h$,
the solutions approach the corresponding regular gravitating solution
on the interval $]r_h,\infty[$ when the limit 
$r_h \rightarrow 0$ is considered.
It is therefore a natural question to investigate how nutty-dyons behave
in the same limit. 

Considering this problem for a few generic values of $(\alpha,n)$  
we reach the conclusion that, in the limit $r_h \rightarrow 0$, the nutty dyon
becomes singular at $r=0$ because the value $\sigma(r_h)$ tends to zero.
 This situation is illustrated on Figure 4 where the functions
$N(r), \sigma(r)$ and $u(r)$ are plotted for three different values
of $r_h$  and $\alpha=1,~n=0.5$. 

Remarkably, this Figure reveals that the functions $\sigma(r)$ and $u(r)$ 
are rather independant of $r_h$ (it is also true for $w(r),\phi(r)$ 
which are not
represented) while the function $N(r)$ indeed involves 
non trivially with $r_h$. 
Note also that for the metric gauge choice $P(r)=\sqrt{r^2+n^2}$, 
the area of two-sphere $d \Omega^2=P^2(r) (d \theta^2+\sin^2 \theta d\varphi^2)$ 
does not vanish at $r=0$.
However, by choosing a Schwarzschild gauge choice
$P(r)=r$, a straightforward analysis of the corresponding  field 
equations (which can easily be derived from (\ref{action})) implies that it is
not possible to take a consistent set 
of boundary conditions at $r=0$
without introducing a curvature singularity at that point. 
Therefore, no globally regular EYMH solutions are found for $n \neq 0$.

The determination of the domain of nutty dyons for
fixed $(\alpha,n,Q_e)$ and increasing the horizon radius $r_h$ is
very likely an involved problem. Already in the case $n=0$,
discussed in  \cite{Brihaye:1999nn} the numerical analysis
reveals several (up to three) branches of solutions on some
definite intervals of the parameter $r_h$. We believe that
similar patterns could occur for $n>0$ but their analysis
is out of the scope of this letter.

\section{Further remarks}

 In this work we have analysed the basic properties of gravitating YMH system in the
presence of a NUT charge.
We have found that despite the existence of a number of similarities to the $n=0$ case
(for example the presence of a maximal value of the coupling constant $\alpha$),
the NUT-charged solutions exhibits some new qualitative features.
 
 The static nature of a
$n=0$ spherically symmetric gravitating nonabelian solution implies that
it can only produce a "gravitoelectric" field. 
There both nonabelian monopole and dyon black hole solutions are possible to exist, 
with a well defined zero event horizon radius limit. 
For a nozero NUT charge, 
the existence of the cross metric term $g_{\varphi t}$ shows
that the solutions have also a "gravitomagnetic" field. 
The term $g_{\varphi t}$ does not produce
an ergoregion but it will induce an effect similar to the dragging
of inertial frames \cite{Zimmerman:kv}.
In this case
we have found that only nonabelian dyons are possible to exist and 
the usual magnetic charge quantization
relation is lost.
The total mass of these solutions may be negative and 
the configurations do not survive in the limit of zero event horizon radius.

A discussion of possible generalizations of this work should start with
the radially excited nutty dyons, for which the gauge function $\omega(r)$ possesses nodes.
These configurations are very likely to exist, 
continuing for $n>0$ the excited configurations discussed in \cite{Brihaye:1999nn}.
Also, in our analysis, to simplify the general picture, we set the Higgs potential $V(\phi)$ to zero.
We expect to find the same qualitative results for 
a nonvanishing scalar potential (at least if the parameters
are not to large).
It would be a challenge to construct axially symmetric NUT-charged dyons (the corresponding 
$n=0$ monopoles are discussed in \cite{Hartmann:2001ic}).
Such dyon solutions would present a nonvanishing angular momentum, generalizing the
abelian Kerr-Newman-NUT configurations
(a set of asymptotically flat rotating solutions have been 
considered recently in \cite{Kleihaus:2004gm}). 

Similar to the case $n=0$, the solutions discussed in this
work can also be generalized by including a more general matter
content. However, we
expect that these more general configurations will present the same generic
properties discussed in this work. 
This may be important, since
 there are many indications
that the NUT charge is an important ingredient in low energy string
theory \cite{Johnson:1994ek}, conclusion enhanced by the discovery
of "duality" transformations which relate superficially very
different configurations. In many situations, if the NUT charge is
not included in the study, some symmetries of the system remain
unnoticed (see e.g. \cite{Alonso-Alberca:2000cs} for such an example). 
Therefore, we may
expect the NUT charge to play a crucial role in the duality properties
of a (supersymmetric-) theory presenting gravitating nonabelian dyons.

Unfortunately, the pathology of closed timelike curves is not
special to the vacuum Taub-NUT solution but afflicts all 
solutions of Einstein equations
solutions with "dual" mass in general \cite{Magnon}.  This
condition emerges only from the asymptotic form of the fields, and
is completely insensitive to the precise details of the nature of
the source, or the precise nature of the theory of gravity at short
distances where general relativity may be expected to break down
\cite{Mueller:1986ij}. 
This acausal behavior precludes the nutty dyons solutions discussed in this paper
having a role classically and implies a number of pathological 
properties of these configurations.

Nevertheless, there are various
features suggesting that the Euclidean version
of NUT-charged solutions play an important role in 
quantum gravity \cite{Hawking:ig}.
For example, the entropy of such solutions generically do not obey the simple
"quarter-area law". 
As usual, a positive-definite metric is found by
considering in (\ref{metric}) the analytical continuation 
$t\rightarrow it ,~~n\rightarrow i~n$,  which gives $P^2(r)=r^2-n^2$.
In this case, the absence of conical singularities at the root $%
r_{h}$ of the function $N(r)$ imposes a periodicity 
in the Euclidean time coordinate 
\begin{equation}
 \label{new-rel}
\beta =\frac{4\pi }{N^{\prime }(r_{h})\sigma(r_h)}, 
\end{equation}%
which should be equal with the one to remove the 
Dirac string $\beta=8 \pi n$.
In the usual approach, the solution's parameters 
must be restricted such that the fixed
point set of the Killing vector $\partial _{t}$ is regular at the
radial position $r=r_{h}$. We find in this
way two types of regular solutions, "bolts" (with arbitrary $r_{h}=r_{b}>n$) or
"nuts" $(r_{h}=n)$, depending on whether the fixed point set is of
dimension two or zero (see   \cite{Hawking:1998jf} for a discussion
of these solutions in the vacuum case and  \cite{Astefanesei:2004kn}
for a recent generalization with anti-de Sitter asymptotics).

We expect that the Euclidean nutty dyons will present some new features
as compared to the Lorentzian counterparts. 
For example, globally regular solutions
may exist in this case, since
$r=r_h$ corresponds to the origin of the coordinate system
(note also that the SU(2) Yang-Mills system is known to present 
self-dual solutions in the background 
of a vacuum Taub-NUT instanton \cite{Pope:1978kj}).
In the absence of closed form solutions, the properties of these non-self dual 
EYMH solutions cannot be 
predicted directly from those of the Lorentzian configurations.
However, similar to the Lorentzian case, they can be studied in a systematic way, 
by using both analytical and numerical arguments.
For example, 
the magnitude of the electric potential at infinity of the Euclidean solutions,
is not restricted. Also, the condition $\beta=8\pi n$ implies
$N^{\prime }(r_{h})\sigma(r_h)=2n$ and introduces a supplementary constraint
on the matter functions as $r \to r_h$.
A study of such solutions may be important in a quantum gravity context.
\\
\\
\\
\medskip
\noindent
{\bf\large Acknowledgements} \\
ER thanks D. H. Tchrakian for useful discussions.
YB is grateful to the
Belgian FNRS for financial support.
The work of ER is carried out
in the framework of Enterprise--Ireland Basic Science Research Project
SC/2003/390 of Enterprise-Ireland.



\begin{thebibliography}{99}

\bibitem{'tHooft:1974qc}
G.~'t Hooft,
Nucl.\ Phys.\ B {\bf 79} (1974) 276;
\\
A.~M.~Polyakov,
JETP Lett.\  {\bf 20} (1974) 194 [Pisma Zh.\ Eksp.\ Teor.\ Fiz.\
{\bf 20} (1974) 430].
\bibitem{Julia:1975ff}
B.~Julia and A.~Zee,
Phys.\ Rev.\ D {\bf 11} (1975) 2227.

\bibitem{Volkov:1999cc}
M.~S.~Volkov and D.~V.~Gal'tsov,
Phys.\ Rept.\  {\bf 319} (1999) 1;
\newline
D.~V.~Gal'tsov,
``Gravitating lumps,''
arXiv:hep-th/0112038.
\bibitem{Brihaye:1999nn}
Y.~Brihaye, B.~Hartmann, J.~Kunz and N.~Tell,
Phys.\ Rev.\ D {\bf 60} (1999) 104016 [arXiv:hep-th/9904065];
%
\bibitem{Breitenlohner:1991aa}
P.~Breitenlohner, P.~Forgacs and D.~Maison,
Nucl.\ Phys.\ B {\bf 383} (1992) 357.
\bibitem{NUT}
E. T. Newman, L.Tamburino and T. Unti, J. Math. Phys. {\bf 4} (1963)
915.
\bibitem{Misner}
C. W. Misner, J. Math. Phys. {\bf 4} (1963) 924;
\\
C. W. Misner and A. H. Taub, Sov. Phys. JETP {\bf 28} (1969) 122.
\bibitem{misner-book}
C. W. Misner, in \emph{Relativity Theory and Astrophysics I:
Relativity and Cosmology}, edited by J. Ehlers, Lectures in Applied
Mathematics, Volume 8 (American Mathematical Society, Providence,
RI, 1967), p. 160.
\bibitem{Hawking}
S.W. Hawking, G. F. R. Ellis, \emph{The large structure of
space-time}, Cambridge, Cambridge University Press, (1973).
\bibitem{Lynden-Bell:1996xj}
D.~Lynden-Bell and M.~Nouri-Zonoz,
Rev.\ Mod.\ Phys.\  {\bf 70} (1998) 427
[arXiv:gr-qc/9612049].

\bibitem{dam}
M. Damianski and E. T. Newman, Bull. Acad. Pol. Sci. {\bf 14} (1966)
653;
\\
J. S. Dowker, Gen.\ Rel.\ Grav.\  {\bf 5} (1974) 603.
\bibitem{Gibbons:1986df}
G.~W.~Gibbons and N.~S.~Manton,
Nucl.\ Phys.\ B {\bf 274} (1986) 183.
\bibitem{Hawking:1998jf}
  S.~W.~Hawking and C.~J.~Hunter,
  Phys.\ Rev.\ D {\bf 59} (1999) 044025
  [arXiv:hep-th/9808085].

\bibitem{Mann:1999pc}
  R.~B.~Mann,
  Phys.\ Rev.\ D {\bf 60} (1999) 104047
  [arXiv:hep-th/9903229].

\bibitem{Gross:1983hb}
  D.~J.~Gross and M.~J.~Perry,
  Nucl.\ Phys.\ B {\bf 226} (1983) 29.
\bibitem{Sorkin:1983ns}
  R.~D.~Sorkin,
  Phys.\ Rev.\ Lett.\  {\bf 51} (1983) 87.
\bibitem{Brill}
D.~R.~Brill, Phys.\ Rev. {\bf 133} (1964) B845.
\bibitem{Johnson:2004zq}
C.~V.~Johnson and H.~G.~Svendsen,
Phys.\ Rev.\ D {\bf 70} (2004) 126011
[arXiv:hep-th/0405141].

\bibitem{Radu:2002hf} E.~Radu, 
Phys.\ Rev.\ D \textbf{67} (2003) 084030
[arXiv:hep-th/0211120].

\bibitem{89}
M.~S.~Volkov and D.~V.~Galtsov,
JETP Lett.\  {\bf 50} (1989) 346;
\\
H.~P.~Kuenzle and A.~K.~Masood- ul- Alam,
J.\ Math.\ Phys.\  {\bf 31} (1990) 928;
\\
P.~Bizon,
Phys.\ Rev.\ Lett.\  {\bf 64} (1990) 2844.
\bibitem{bizon}
P. Bizon, O.T. Popp, Class. Quant. Grav. {\bf 9} (1992) 193;
\\
A.A. Ershov, D.V. Galtsov, Phys. Lett.  A {\bf150} (1990) 159;
\\
D.~V.~Galtsov and A.~A.~Ershov,
Phys.\ Lett.\ A {\bf 138} (1989) 160.
\bibitem{Hawking:ig}
S.~W.~Hawking in {\it General Relativity. An Einstein Centenary
Survey}, edited by S.~W.~Hawking and W.~Israel, (Cambridge,
Cambridge University Press, 1979) p. 746.
\bibitem{Mueller:1986ij}
M.~Mueller and M.~J.~Perry,
Class.\ Quant.\ Grav.\  {\bf 3} (1986) 65.
\bibitem{Forgacs:1980zs}
P.~Forgacs and N.~S.~Manton,
Commun.\ Math.\ Phys.\ {\bf 72} (1980) 15.
\bibitem{Bergmann:fi}
P.~G.~Bergmann and E.~J.~Flaherty,
J.\ Math.\ Phys.\  {\bf 19} (1978) 212.
\bibitem{vanderBij:2002sq}
J.~J.~van der Bij and E.~Radu,
Int.\ J.\ Mod.\ Phys.\ A {\bf 18} (2003) 2379
[arXiv:hep-th/0210185].
\bibitem{Johnson:1994ek}
C.~V.~Johnson and R.~C.~Myers,
Phys.\ Rev.\ D {\bf 50} (1994) 6512
[arXiv:hep-th/9406069].
\bibitem{Magnon}
S.~Ramaswamy and A. Sen, J.\ Math.\ Phys.\  {\bf 22} (1981) 2612;
\\
A.~Magnon, J.\ Math.\ Phys.\  {\bf 27} (1986) 1066;
\\
A.~Magnon, J.\ Math.\ Phys.\  {\bf 28} (1987) 2149.
\bibitem{COLSYS}
 U. Ascher, J. Christiansen, R.~D. Russell,
 Math. of Comp. {\bf 33} (1979) 659;\\
 U. Ascher, J. Christiansen, R.~D. Russell,
 ACM Trans. {\bf 7} (1981) 209.

\bibitem{Zimmerman:kv}
R.~L.~Zimmerman and B.~Y.~Shahir,
Gen.\ Rel.\ Grav.\  {\bf 21} (1989) 821.
\bibitem{Hartmann:2001ic}
B.~Hartmann, B.~Kleihaus and J.~Kunz,
Phys.\ Rev.\ D {\bf 65} (2002) 024027
[arXiv:hep-th/0108129].
\bibitem{Kleihaus:2004gm}
B.~Kleihaus, J.~Kunz and F.~Navarro-Lerida,
Phys.\ Lett.\ B {\bf 599} (2004) 294
[arXiv:gr-qc/0406094].
\bibitem{Alonso-Alberca:2000cs}
N.~Alonso-Alberca, P.~Meessen and T.~Ortin,
Class.\ Quant.\ Grav.\  {\bf 17} (2000) 2783
[arXiv:hep-th/0003071].

\bibitem{Astefanesei:2004kn}
  D.~Astefanesei, R.~B.~Mann and E.~Radu,
  JHEP {\bf 0501} (2005) 049
  [arXiv:hep-th/0407110].
\bibitem{Pope:1978kj}
  C.~N.~Pope and A.~L.~Yuille,
  Phys.\ Lett.\ B {\bf 78} (1978) 424.


\end{thebibliography}
\end{document}